# On the physical and circuit-theoretic significance of the Memristor

Emanuel Gluskin

*Abstract*—It is noticed that the inductive and capacitive features of the memristor reflect (and are a quintessence of) such features of *any resistor*. The very presence in the resistive characteristic $v = f(i)$ of the voltage and current state variables, *associated, by their electrodynamics sense, with electrical and magnetic field*s, forces any resister to cause to accumulate some magnetic and electrostatic fields and energies around itself, i.e. *L* and *C* elements are always present. From the *circuit-theoretic point of view*, the role of the memristor is seen, first of all, in the eliminating of the use of a unique *v*(*i*). This makes circuits with hysteretic characteristics relevant, and suggests that the concept of memristor should influence the basic (general) problem of definition of nonlinearity. Since memristor mainly originates from resistor, it was found methodologically necessary to overview some unusual cases of resistive circuits.

*Index Terms*—Resistor, Memristor, Electromagnetic Fields, Energy flow, Circuit Theory, Characteristic, Hysteresis, Education.

## I. INTRODUCTION

Predicted by Professor (Academician) Leon O. Chua in [1], and already technologically realized (as a two-state element, a switch) [2], the *memristor* (i.e. "resistor with memory") and "memristive circuits" became a popular research topic, e.g. [3-6] and references there. The original purpose of this specific circuit element is to connect electrical charge with magnetic flux, which is, as we stress after [1,5], outside the borders of the use of the resistive "characteristic" given by a single equality $v = f(i)$, with a unique $f(.)$.

We have to start, however, (improving [7]) with an argument showing that the electrical and magnetic fields, -- which are the organic features of the *C* and *L* elements, -- are accumulated in the space around any conductive element. These fields are just *necessary* for the power losses in the resistor even in the dc process.

In this view, the memristor, in which the *L*- and *C*- features "officially" (in a circuit context [1,3-6]) appear together with the *R*-feature, becomes a "representative" of any existing resistor, which allows one to better understand both the theoretical role of the memristor and that of the usual resistor. The latter is relevant, for instance, to consideration of the frequency limits [8,9] for applications of Kirchhoff's equations.

On the main line of "escaping" the use of resistive "characteristics" we then consider in detail the example of the strongly nonlinear and practically very important fluorescent lamp circuits. Finally, we overview some power-law resistive circuits allowing one to create interesting hysteresis characteristics.

## II. THE USUAL RESISTOR AND THE ENERGY FLOW

Circuit theory [8,9] defines the resistor as a 1-port described by the voltage-current (*v-i*) *one-to-one* relation

$$v = f(i) , \qquad (1)$$

while practical engineers often tend to understand the term "resistor" *only* as related to the linear dependence $v = Ri$ where the constant coefficient *R* is the *resistance.*

An immediate observation (unfortunately, does not appearing in textbooks) is that since (1) is formulated in terms of the capacitor's and inductor's *state variables* (*v* for *C*, and *i* for *L*), *no resistor can exist without some associated capacitance and inductance.*

Indeed, for any flowing current *i* (may be a *direct current*), there is magnetic field *H* around the wire (the element), and thus there is the magnetic energy ~ $H^2$ associated with some inductance. Similarly, if we have a voltage drop *v* on an element, then we have electrical field *E* of which *v* is an integral measure, and since the tangential *E* is continuous on the surface of the conductor [10], it exists also outside the conductor, and we have electrostatic energy ~ $E^2$ accumulated around, as that of a capacitor. (Of course, the fields and the energies exist also inside the conductor.)

Furthermore, since the current flows in the direction of *E*, *E* appears to be perpendicular to *H* around the conductor, so that the nonzero Pointing-vector (the vector product) *S* = [*E*,*H*] is directed towards the conductor, and the energy comes from the outer space to the element; in this way it is dissipated in the resistor (i.e. is heating it) as instantaneous power $p = vi$ (for the linear case, $Ri^2$) supplied to the resistor.

Remark 1: For cylindrical form of the conductor, the direct proof is very simple. The electrical power flowing into the conductor is *sS* where *s* is the relevant conductor's surface. Using also the length *l* of the cylinder and its radius *r*, we have for the flow of the energy into the conductor

$$sS = sEH = (l 2\pi r)(\frac{v}{l})(\frac{i}{2\pi r}) = vi \qquad (2)$$

i.e. precisely the consumed power *p*. Notice that the dependence of *v* on *i* is not important, the conductor can be electrically nonlinear in any degree.

Thus, even the state with the established *dc* current, is, essentially, a *field problem* [11] in which the surface of the

Submitted 18 Sept. 2014.
E. Gluskin is with Kinneret College in the Jordan Valley (on the Sea of Galilee), Tzemach, 15132 Israel (email: gluskin@ee.bgu.ac.il).

resistor plays the role of the boundary of the space with the field. Figure 1 (used in [7]) illustrates the point.

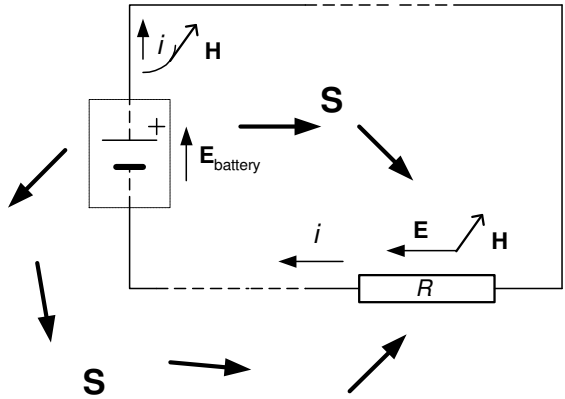

Fig. 1: The energy (power) enters the resistor from outside because of the fields **E** and **H**, and thus $p = p_R = Ri^2$ (or, more generally, $vi$) is obtained. The simplest dc state is in the focus. See also the references in [7], in which the field analysis is presented in detail and for any frequency. One can see here, in particular, a Thevenin equivalent.

*Thus, by the very sense of the variables v and i defining the usual resistor, there always is some accumulated electromagnetic energy. In other words, there are a capacitor and an inductor organically associated with any usual resistor, in any circuit.*

Of course, one should *not* interpret these absolutely necessary capacitor and inductor as some "parasitic" elements, because no parasitic element defines the *basic properties* of a circuit (here, the value of *p*).

From the positions of field theory, the existence of both electrical and magnetic fields *in the dc state* is associated just with the fact that the boundary is conductive, i.e. there cannot be *v* without *i*, or *E* without *H*. It would be interesting to consider the mutual influence of the circuit and field problems in more detail; for instance, the cases of superconductive conductors (when the battery becomes a current-source), several parallel conductors, etc. .

### III. THE FIELDS AS A REQUIREMENT OF THE THEORY OF RELATIVITY

Regarding the physical situation illustrated by Fig. 1, it should be advised to start the electromagnetic field analysis of circuit elements, by drawing one's attention to the fact that if we apply Ohm's law to a series connection of so many resistors that the length of the connection is many kilometers, -- then, applying the battery to the chain and assuming that the *momental (instantaneous)* relation $v = Ri$ is correct, we can *immediately* transfer the *signal as the current* (i.e. the information about connecting the source) over the whole distance, *which obviously contradicts the theory of relativity.*

It thus becomes clear that after closing the switch there is an initial electromagnetic wave, propagating from the battery and leading to a finally established process with the dc current, while at this very dc state *there is a steady flow of energy from the battery to the resistor via the outer space*, associated with an inductance and a capacitor. $\vec{S} \neq \vec{0}$!

### IV. ON THE MATHEMATICAL (CIRCUIT DESCRIPTION) SIDE

That in the description of the memristor no "characteristics" $v(i)$ is directly used, is of crucial importance. In [5], the following system of equations is used for a memristive 1-port:

$$v = R(\vec{x},i)i$$
$$\frac{d\vec{x}}{dt} = \vec{f}(\vec{x},i) \qquad (3)$$

where $\vec{x}$ is the vector of the internal state variables of the memristor. The memory is included because of the derivative in the latter equation; solutions of differential equations depend on the initial conditions, i.e. on the past.

The appearance in (3) of the scalar function *R* (the "memristance") depending on *several* variables is an interesting point, because in order to come from a usual resistor to a memristor, *one* such variable is sufficient. Work [12] considers reduction (via some functions' "superposition") of the number of the arguments of a function of many variables to functions of a smaller number of variables. Since from a system of first order equations for several (many) variables (the second line of (3)), one can derive one equation of a higher order for one variable, it is interesting to know when $R(.,i)$ can include only one such variable.

In order to clarify how system (3) reflects the basic definition [1] of an element which mutually connects magnetic flux

$$\psi(t) = \int_t v(t)dt$$

and electrical charge

$$q(t) = \int_t i(t)dt ,$$

we observe that from the second equation of (3), $\vec{x}(t)$ should depend on $\int_t i(t)dt = q(t)$. Then, from the first equation,

$$v(t) = F(q(t),i)i$$

with some function $F(.,.)$, and

$$\psi(t) = \int_t v(t)dt = \int_t F(q(t),i)i\,dt = \int_q F(q,i)dq .$$

Thus, if $i(t) = dq/dt$, included in *F*, can be directly expressed via *q* (that is, some differential equation for $q(t)$ can be derived), then we have a direct connection of $\psi$ with $q$. Thus, according to model (3), the *charge variable* has to be well defined.

We can, however, pass on to the "dual" situation of:

$$i = G(\vec{x},v)v$$
$$\frac{d\vec{x}}{dt} = \vec{f}(\vec{x},v)$$

Then, having $\vec{x}(t)$ dependent on $\int_t v(t)dt = \psi(t)$, we have from the first equation that $i(t) = F(\psi(t),v)v$, and

$$q(t) = \int_t i(t)\,dt = \int_t F(\psi,v)v\,dt = \int_\psi F(\psi,\frac{d\psi}{dt})d\psi.$$

In this case, in order to connect $q$ and $\psi$, a good description of the *flux variable* (i.e. a differential equation for $\psi(t)$) is needed.

As we further note in Section X, model (3) can be seen as problematic in an energetic sense.

A more simple question re (3) is that of the possibility of $R(.,.)$ to directly include $di/dt$. This question interests us in view of [13] where the $v(t)$-$i(t)$ dependence for the practically very important fluorescent lamp appears as

$$v = f(i, di/dt)i. \qquad (4)$$

We observe re (4) that it can *not* be written as any $v(i)$, and that the (weak) inductive feature of the lamp is obvious from the dependence of $v$ on $di/dt$.

Figs 2 and 3 show two hysteresis characteristics of real fluorescent lamps, which we model using sign[$i$]). Since for the fluorescent lamp circuit, the averaged consumed power, $P = <p(t)>$, is the main parameter, it is important for a model to give correct values of $v$ at the *high* values of $i$, when $p(t)$ is relatively large. Thus, function "signum" is appropriate here.

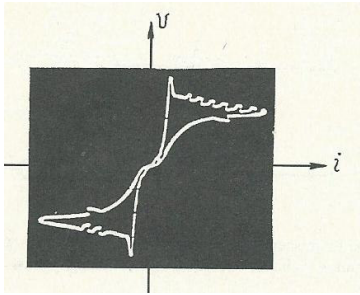

Fig 2: The (mainly) resistive hysteresis loop of a fluorescent lamp. The direction of the passing the loops is clockwise; the voltage obtains its highest value *before* the current, which is an inductive feature.

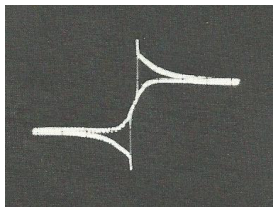

Fig 3: Such a loop for a fluorescent lamp with smaller diameter of the tube, i.e. smaller mass of the gas.

The *inductive* feature of the lamp is associated with the *clockwise* direction of its hysteresis $v$-$i$ loop. (Indeed, *for a pure inductor*, having the Lissajous figure in the $i$-$v$ plane as a circle, the voltage is leading the current, i.e. the point ($i = 0$; $v = v_{max}$) comes in time before the point ($i = i_{max}$; $v = 0$).) The opposite direction of such a loop means capacitive behavior. See [13] for more details and see also [14] in which hysteresis characteristics are discussed.

One observes that any hystersis characteristic very simply relates to memory, because movement along the hysteresis loop in any direction is associated with *remembering* where we previously were on the loop.

Another remarkable fact, stressed in [13], is that the energy associated with the lamp's inductance $L'$,

$$W_{L'} = L'\frac{i^2}{2}, \qquad (5)$$

is of an *electrostatic origin*. This energy is associated with the separation of positive and negative charges because of the "bipolar" diffusion. There is no any significant magnetic flux associated with the lamp to justify the rather large value of $W_{L'}$. *Thus, introducing di/dt into (4) we mathematically obtained an inductor, but its energy appears to be of a capacitor's nature. In diffusion processes, the current can be delayed with respect to the voltage, which is an inductive feature per se, agreeing with v ~ di/dt, or i be an integral of v, but not having any relation to any significant magnetic flux and magnetic energy.* See also [15,16], concerned with electrical features of human body.

It should be also noted, to our main point, that already for the (purely resistive) $v(i)$-model of first approximation of the lamp,

$$v(i) = A\, sign[i], \qquad (6)$$

with a constant $A$, it is natural to try to transfer from this *characteristic* $v(i)$ to the dependence $v(t)$-$i(t)$ between the *time functions*:

$$v(t) = A\, sign[i(t)], \qquad (7)$$

allowing the periodicity of $i(t)$ to be well used in the Fourier analysis of the circuit, since for $i(t)$ without pauses, sign[$i(t)$] is a simple rectangular wave [13,17]. That is, *singularity* of a characteristic can change (define) its descriptional role. That for many *switched* circuits and systems no "characteristic" (not only of the simple hardlimiter type) exists is explained in [18], essentially using the fact that the relations *between time-functions* are dominant.

### V. AN EXAMPLE OF MODELING OF ELEMENTS WITH HYSTERESIS CHARACTERISTICS

The following modeling further demonstrates the transfer from a given $v(i)$ to some $v(t)$-$i(t)$. This transfer is done in some *analytically-iterative* way. Using initially the purely hardlimiter model (6), we then take the hysteresis into account by both adding the inductive term *and increasing the factor before* sign[$i$] (*"A"* remains with its sense as $v$ at $i_{max}$) which is required by power reasons, obtaining

$$v(t) = A_1 sign[i(t)] + L' \frac{di}{dt} \quad . \quad (8)$$

Since both of the inequalities

$$L' > 0 \quad ; \quad A_1 > A$$

exist because of the same hysteresis phenomenon, the difference $A'-A$ and $L'$ must be analytically connected,

$$A_1 - A \sim L',$$

and because of the dimensional reasons,

$$\frac{A_1 - A}{A} = k \frac{L'}{L},$$

where $L$ is the inductance of the ballast shown in Fig. 4, and $k$ is a non-dimensional constant ($k = 2$ is obtained below). Thus,

$$A_1 = A(1 + k\frac{L'}{L}).$$

The presence of the external inductance $L$ in eq' (8), written now as

$$v(t) = A(1 + k\frac{L'}{L})sign[i(t)] + L'\frac{di}{dt}, \quad (9)$$

shows that it is better to define the lamp not independently, but in the circuit context. This methodologically important point will be immediately treated in detail.

A systematic nonlinear theory of fluorescent lamp circuits based on the simplest model (6) is given in [17]; the derivation of (9) is presented in [19]. The main steps of this derivation are as follows.

Regarding the circuit of Fig. 4,

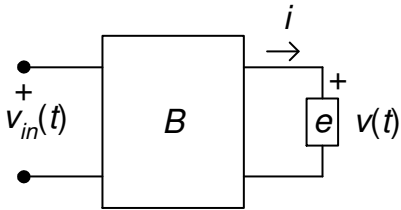

Fig.4: The lamp '$e$' is connected via linear "ballast" $B$ including a big inductance $L$ that provides uninterrupted current $i(t)$. The ballast can be series; then $\hat{L}_1 = \hat{L}_2$ in (10), but the presentation $i = i_1 + i_2$ remains.

we write

$$\begin{aligned} i(t) &= \hat{L}_1[v_{in}(t)] - \hat{L}_2[v(t)] \\ &= \hat{L}_1[v_{in}(t)] - A\hat{L}_2[sign[i(t)]] \end{aligned} \quad (10)$$

where $\hat{L}_1$ and $\hat{L}_2$ are some *linear integral operators* representing the steady state output-current response of $B$ to its inputs. The last term of (10) shows that we start with model (6) for the lamp. However, (10) will be immediately used for correcting this model. Let us come to (9) via

$$v(t) = A\,sign[i(t)] + L_1\frac{di_1}{dt} - L_2\frac{di_2}{dt} \quad (11)$$

introducing *two* inductive parameters $L_1$ and $L_2$, and splitting $i(t)$ into two parts, $i_1(t)$ and $i_2(t)$, such that $i_2(t)$ is more singular, i.e. its Fourier coefficients are $O(1/n^2)$, while Fourier coefficients of $i_1(t)$ are $O(1/n^m)$, $m > 2$. In order to separate the current components, we rewrite (10) as

$$\frac{di}{dt} = \frac{d}{dt}\hat{L}_1 v_{in}(t) - A\left(\frac{d}{dt}\hat{L}_2 sign[i(t)] - \alpha sign[i(t)]\right) - \alpha A sign[i(t)]$$

and find $\alpha$ so that the parenthetic term is the "smooth" (continuous) one, i.e. relates to $di_1/dt$. Thus, for such $\alpha$, for the more singular $i_2$, we have

$$\frac{di_2}{dt} = -\alpha A\,sign[i(t)] \quad . \quad (12)$$

It appears [19] that $\alpha$ is the converse value of the *asymptotic inductance* $L$ of $B$, that is

$$\alpha = \frac{1}{L} \equiv \lim_{n \to \infty} n\omega |Y(n\omega)|,$$

where $Y(.)$ is the *admittance function* [8,9] of $B$.

Substituting into (11) first

$$\frac{di_1}{dt} = \frac{di}{dt} - \frac{di_2}{dt}$$

and then using (12), we finally obtain (11) as

$$v(t) = A(1 + \frac{L_1 + L_2}{L})sign[i(t)] + L_1\frac{di}{dt}.$$

Setting for simplicity $L_1 = L_2$ and denoting this common value as $L'$, we obtain (9) as

$$v(t) = A(1 + 2\frac{L'}{L})sign[i(t)] + L'\frac{di}{dt}.$$

This (hysteresis) model is used in [13], well explaining the main features of the real lamp circuits.

One notes that for any *linear* model of the lamp, it is unclear why the ballast is at all needed, and the actually problematic high harmonic currents can not be estimated. Unfortunately, the theory of [17,19,13], never included in any textbook, is generally unknown to the power-system engineers, and the absolutely unsatisfactorily linear *R-L* model of the lamp is still used. (The high harmonic currents are not calculated, they are measured by the electrical company, and then one pays the fine if these currents are too high.)

## VI. THE "ZEROCROSSING NONLINEARITY" AVOIDING THE V(I) CHARACTERISTICS

We interpret the original purpose of [1] to find fourth basic circuit element connecting electrical charge with magnetic flux, as an attempt to eliminate the use of direct "resistive" *v*(*i*)-relations. This seemingly somewhat narrow interpretation is a fruitful one, and we shall continue this line, by assuming that any way of analyzing circuits including resistors, without usage of direct *v*(*i*)-relations, can be relevant to memristive circuits. This hypothesis should open a wide field of "memristive switched circuits".

Let us proceed with noting that the already considered nonlinear theory of fluorescent lamp circuits can be formulated using the concept of *zerocrossing nonlinearity*. This concept, very important for us now, generally means [20] that zerocrossings $\{t_k\}$ of a time function related to the system under study *cannot be prescribed*.

Since the zerocrossings are some analytical parameters, let us compare them to any other parameters that can appear in the equations that describe the system. Thus, when speaking about coefficients of a differential equation, we can separate between three cases:

a. *these coefficients are constant*

b. *they are given as some known time functions*

c. *they depend on the state variables to be found.* In this case, strictly speaking, they are already not any "coefficients", but it is *useful* to see them on their "original places", as some "structural parameters", as if the dependence on the state variables appears gradually, see [20] and below.

In the cases "a" and "b" (of, respectively, LTI and LTV systems) the coefficients are prescribed a priori, and in the case "c" (of nonlinear systems) they cannot be prescribed.

The situation with $\{t_k\}$ is absolutely the same, and that $\{t_k\}$ cannot be prescribed means a nonlinearity, namely, "zerocrossing nonlinearity".

Regarding the explicit introduction of the zerocrossings into the mathematical description of a system, we note that when an uninterrupted (i.e. not having intervals where it would be identically zero) zerocrossing periodic current function *i*(*t*) is provided, the nonlinear relation (6) appears in the theory of the lamp circuits as a square wave time function

$$sign[i(t)] = \frac{4}{\pi} \sum_{1,3,5,...}^{\infty} \frac{\sin[n\omega(t-t_1)]}{n} \quad (13)$$

where $t_1$ (modulo *T*, $T = 2\pi/\omega$) is a -/+ zerocrossing of *i*(*t*).

With this observation, return to Fig. 4 and take

$$v_{in}(t) = U\xi(t)$$

where *U* is a scaling parameter, and $\xi(t)$ is a *T*-periodic wave that defines the waveform of the input function and the position of the time origin, i.e. the point *t* = 0. Changing *U*, we can observe nonlinear effects in this driven circuit. For instance, observe that its *steady-state* power consumption is *not* proportional to $U^2$.

It is not unexpected that some nonlinearity is thus easily observed, but this nonlinearity can take, classificationally, two very different forms, respectively to whether the zerocrossings of *i*(*t*), are dependent on *U* (i.e. are moved with respect to a zero of $\xi(t)$ ), which is the usual case, or are independent of *U* (i.e. unmoved, constant) which can be provided, in some range of *U*, by a special syntheses of *B* [17,13].

In order to illustrate the classification aspect, consider a simple version of the circuit, which is described by the equation

$$L\frac{di}{dt} + A sign[i(t)] + \frac{1}{C}\int^t i(\lambda)d\lambda = v_{in}(t), \quad (14)$$

i.e.

$$\left(L\frac{di}{dt} + \frac{1}{C}\int^t i(\lambda)d\lambda\right) + \left(\frac{4A}{\pi}\sum_{1,3,5,...}^{\infty}\frac{\sin[n\omega(t-t_1)]}{n}\right) \quad (15)$$
$$= U\xi(t).$$

If $t_1$ depends on *U*, then the second parenthetic term in (15) is obviously nonlinear (the "zerocrossing nonlinearity").

If $t_1$ is independent of *U*, then this term presents a known time function. However, in this case the system is *also* nonlinear, *in the sense of the affine nonlinearity*. Indeed, denoting the term with the sum as -*f*(*t*), we can rewrite (15) as

$$(\hat{L}i)(t) = U\xi(t) + f(t) ,$$

with a linear operator $\hat{L}$, and it is obvious that $U \to kU$ does *not* lead to $i \to ki$.

Thus, in agreement with the fact that the fluorescent lamp that is present in the circuit is nonlinear disregarding the possible specific synthesis of *B* providing $t_k(U)$ = const, it appears that the map $U \to i$ in (15) is always nonlinear. However, the appearance of the case of affine nonlinearity is an interesting specificity of the zerocrossing nonlinearity, and one sees that the (seemingly formal) affine nonlinearity can be rich in the physical context.

From the positions of circuit theory [8,9], the above use of *f*(*t*) is an application of the "substitution theorem", in which a passive element is replaced by a source, but such an application is very rarely so constructive as here.

## VII. ON THE FORMALIZATION OF THE "ZEROCROSSING NONLINEARITY"

The formalization of the above argument is given by the state equations written in the "structural form", that is, so that the structure would be given by matrices [A] and [B] (let us focus only on [A]) as

$$\frac{d\vec{x}}{dt} = [A(t,\vec{x})]\vec{x} + [B]\vec{u} . \quad (16)$$

Objecting this writing, and accepting only that of

$$\frac{d\vec{x}}{dt} = \vec{F}(t, \vec{x}, \vec{u}), \quad (17)$$

by the argument that when a coefficient becomes dependent on $\vec{x}$ it stops to be a coefficient, means not-understanding the heuristic importance of the concept of structure. The best pedagogical and heuristic definition of nonlinearity of a system is that the system's structure is dependent on $\vec{x}$, or on the inputs. Thus, for instance, the nonlinearity of (still unwritten) hydrodynamic equations is directly seen from the fact that the velocity field of a liquid flow is *both* the variable to be found and the "structure" of the flow (system), which is the "philosophy" of (16) and not of (17).

In order to come from this point to the zerocrossing nonlinearity, it remains to observe that for any switched system, LTV or NL, we have

$$\frac{d\vec{x}}{dt} = [A(t, \{t_k\})]\vec{x} + [B]\vec{u}, \quad (18)$$

and if $\{t_k\} = \{t_k(\vec{x})\}$, then we have the NL case of (16).

The following lucid example of the zero-crossing nonlinearity, strongly supports the "philosophy" of (18) and (16). Whether or not such a system may be regarded as a memristive system, depends on the correctness of our research hypothesis that avoiding definitions of nonlinearity by means of "characteristics" is typical for memristive systems.

### VIII. TWO LINEAR (SUB)SYSTEMS CONTROLLED BY A NONLINEAR SWITCHING (THE EXAMPLE OF [21])

Work [21] discusses two topologically similar linear (sub)circuits having (only) one parameter (element) different. The switching from one (sub)circuit to another means that the (topologically similarly defined) outputs of the (sub)circuits are used in turn.

This is, of course, the same as the case of one circuit having (only) one of its parameters (elements) being switched from time to time. That is, we can consider the system of [21] in view of (18). The question is how [21] defines the switching, i.e. whether or not we can transfer here from (18) to (16) having $[A(t, \vec{x})]$.

The switching is defined in [21] by the criterion that a *state variable* crosses some given level. This means $\{t_k\} = \{t_k(\vec{x})\}$, and the system is definitely nonlinear.

That [21] stresses only linearity of the subsystem shows that the zerocrossing (or level-crossing) nonlinearity can be missed even by outstanding circuit specialists.

Contrary to the case of the hardlimiter model, that does not lead [17,13] to any chaos, [21] reports obtaining an interesting chaotic process, though the definition of the moments of switching is done in [21] using computer, which is not "technological".

Thus, using zerocrossing nonlinearity $\{t_k\} = \{t_k(\vec{x})\}$ one can obtain any nonlinear effect. Work [22] also supports this opinion, by demonstrating a chaotic process, obtained by a "reflection from time axes" of a time function at its zerocrossings, in an otherwise linearly-defined circuit.

We continue, with the use of *hysteresis characteristics*, perhaps close to a hardlimiter, now in a very different circuit context.

### IX. HYSTERESIS CHARACTERISTICS USING POWER-LAW CIRCUITS

Among the resistive circuits, a special place belongs to *power-law* circuits [23] (actually, some 1-port grids), composed of similar elements, all with the same *power-law* characteristic

$$v = D_\alpha i^\alpha \quad (19)$$

where the constant $D_\alpha$ has proper dimension. It is sometimes suitable to write (19) using two constants, $v_o$ and $i_o$, having the usual dimensions of volt and ampere:

$$\frac{v}{v_o} = \left(\frac{i}{i_o}\right)^\alpha. \quad (20)$$

Obviously, the point $(i_o, v_o)$ belongs to the $v(i)$-curve for any $\alpha$. (Consider Fig. 5 below.)

The case of $\alpha \to \infty$ is especially interesting here, because then (20) becomes a *current hardlimiter* at the value $i \equiv i_o$:

$$v \underset{\alpha \to \infty}{\to} \begin{cases} 0, & 0 \leq i < i_o \\ \infty, & i_o < i \end{cases}.$$

Since, furthermore, from (20)

$$\frac{i}{i_o} = \left(\frac{v}{v_o}\right)^{1/\alpha},$$

we have

$$i \underset{\alpha \to 0}{\to} \begin{cases} 0, & 0 \leq v < v_o \\ \infty, & v_o < v \end{cases},$$

that is, a *voltage hardlimiter* at the level $v \equiv v_o$, for $\alpha \to 0$.

The initial motivation to consider such power-law elements, generalizing the linear resistor that is obtained here for $\alpha = 1$, is that any 1-port composed of such similar elements also has at its input (i.e. for the input voltage-current relation) a power-law characteristic as in (19) or (20), with the same degree $\alpha$, though having some factor dependent on $\alpha$ and the structure/topology of the 1-port.

The preservation of the degree in the input characteristic allows one to easily calculate the specific fractals [24] obtained from 1-ports using the fact that any branch of a 1-port is also 1-port. (Repeat the given structure of a 1-port in each its branch, obtaining a new, more complicated 1-port, *ad hoc* also as a power-law element.) That the "fractal-nature" of *any* 1-port was never considered in classical circuit theory, is because of absence of the proper analytical tool; the power-law characteristic just provides such a tool.

Works [25,26] report a feature of the power-law circuits, which is named "approximate analytical superposition". This approximate superposition is associated with the generalization of the usual parallel connection of circuits by their inputs to the *connecting node to node*, *for the respective nodes*, the circuits of similar topology. It appears that the input current of such a connection of two different "$\alpha$-circuits", weakly differs from the *sum* of the input currents of each of the $\alpha$-circuit taken separately. The reason for this unexpected and very interesting circumstance is that the nodal voltages of the connection possess some intermediate values with respect to the values of the respective nodal voltages of the separately taken circuits.

As was noted, for $\alpha \gg 1$, we deal with a characteristic that is close to a *current hardlimiter*, and for $\alpha \ll 1$ with a characteristic that is close to a *voltage hardlimiter*.

Using such (electronically realized) characteristics as

$$v = \begin{cases} D_{\alpha_1} |i|^{\alpha_1} \, sign[i], & \dfrac{di}{dt} > 0 \\ D_{\alpha_2} |i|^{\alpha_2} \, sign[i], & \dfrac{di}{dt} < 0 \end{cases} \quad (21)$$

where, generally, $\alpha_2 \neq \alpha_1$, one can try to create memristive grid-type 1-ports, with hysteresis characteristics that for $\alpha_{1(2)} \gg 1$, or $\alpha_{1(2)} \ll 1$, will be close to a hardlimiter characteristic. (In view of [2], to have *one* of the "$\alpha$" to be very large or small is especially interesting for applications.) Fig. 5 illustrates (21). For $\alpha_2 > \alpha_1$ we have inductive loop, and for $\alpha_1 > \alpha_2$, -- a capacitive one.

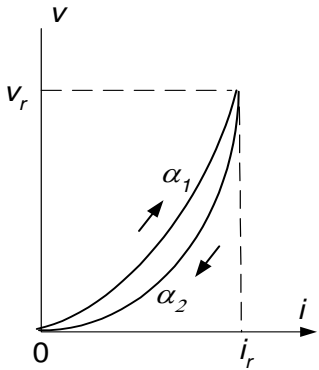

Fig. 5: An illustration to (21), for $\alpha_2 > \alpha_1$ (an inductive loop), presented in the interval $0 < i < i_r$ of the whole interval $-\infty < i < \infty$ where $v(i) = -v(-i)$. The larger is $\alpha_2 / \alpha_1$, the better is the memristor in the sense of [2].

The return point $(i_r, v_r)$ (which is, of course, $(i_o, v_o)$ in the notations of (20)) of this not-smooth "eye-type" loop is found from the equality (we take $i > 0$)

$$D_{\alpha_1} i_r^{\alpha_1} = D_{\alpha_2} i_r^{\alpha_2},$$

i.e.

$$i_r = \left(D_{\alpha_2} / D_{\alpha_1}\right)^{\frac{1}{\alpha_1 - \alpha_2}}, \quad (22)$$

and $v_r$ is then easily found from (21).

If $\alpha_2 = \alpha_1 + 1$, then from (22) $i_r = D_{\alpha_1} / D_{\alpha_2}$.

The case of $v \sim i^{1-\varepsilon \cdot sign[di/dt]}$, $\varepsilon > 0$, (i.e. $\alpha_2 = \alpha_1 + 2\varepsilon$) gives a hysteresis loop that for $\varepsilon \to 0$ becomes a linear resistor characteristic.

## X. CONCLUSIONS AND FINAL REMARKS

The memristor with its *RLC* features is, first of all, a logical "quintessence" of the physical reality of any resistor. Since any use of the variables *v* and *i* introduces a capacitor and an inductor, *any resistor* described by some *v*(*i*) characteristic is associated with energy storing elements. This physical aspect relates to the very foundation of circuit theory, and every circuit specialist has to keep it in mind.

The electrodynamics explanation of the power losses can make the memristor relevant to the theory of large power distribution systems, not only to nano-electronics [2].

In nano-electronics structures, the small size of a local system contributes to a reduction of the internal EM interferences occurring with switching, but for the whole system (as a computer memory) this may be a problem.

Motivated by work [14] with its "intermediate" discussion of some mainly-resistive nonlinear elements (interpreted as memristors) with hysteresis characteristics whose (usually weak) *C* and *L* features are obvious, we have had to overview some nontrivial analytical details of a nonlinear theory of fluorescent lamp circuits, which should be useful for a wider analysis of the elements with hysteresis. These details suggest, in the spirit of the basic approach of [1], avoiding the use of the unique *v*(*i*) characteristics as quickly as possible. This should influence the general problem of definition of circuits' nonlinearities, which is traditionally done via nonlinearity of a "characteristic" of an included element.

Since already for the *purely hardlimiter* model (6), it is reasonable to pass on from *v*(*i*) to *v*(*t*)-*i*(*t*), this model should be related to the class of hysteresis models, -- as the case of zero hysteresis. It follows that if hysteresis relations are typical for memristive devices (such opinion is supported in [14], and in the references there), *then the hardlimiter relation is also relevant to the memristrs*. This is one of the justifications of our attention to "switching nonlinearity" in Sections VI-VII.

It is shown (eq. (9)) that an element can (should) be not necessarily equationally defined when taken separately; rather when it is inside a circuit. Actually, (3) is also in this spirit.

The unusual power-law circuits allow one to create interesting hysteresis characteristics of the "eye" (crescent, lune) type.

In the actual realization of the memristor in [2], its resistance depends on the charge that has passed via the same element. However, from the theoretical point of view (recall *dependent sources* that from the matrix point of view are just some "non-diagonal resistors") the resistance of the element can be dependent on the charge that passed via *another* element of the same 1-port. Indeed, in (3) the initial conditions, associated with the memory, can relate to some *x* that need not be *v* or *i*. Thus a "dependent memristor" can be theoretically obtained, though there is a problem with required activity of the elements; dependent sources have internal sources of energy. Since the memristor is basically defined as a passive element, one can see here a paradox contained in (3).

Undoubtedly, the topic of memristive circuits is a very interesting combination of circuit and physical analysis. Works [14,27,28] are recommended for a completion of the physical side.

From the positions of general system theory, one should try to generalize the conclusions for electrical or electronic systems, to other physical systems. Thus, works [18,20] note that the statistical processes in a gas, i.e. in an ensemble of colliding particles, undergo processes of the type of "switching nonlinearity". That is, the switching nonlinearity is "as old as the world". This should not be forgotten in any attempt to connect the topic of memristors with general dynamic systems.

As is observed in [27] regarding scientific research, every scientific discovery not only opens some new ways for the future, but also allows us to better understand what was done in the past, and it is just methodologically natural that the invention of the memristor causes us to reconsider the usual resistor.

We hope that the present heuristic arguments can contribute to the *logical position* of the memristor in circuit theory, to the analysis of some electromagnetic field problems (Section II), and can encourage a young scientist not only to try to be the first in developing a new application, but also the last in completing an old theory.

Our thinking and our language enrich each other. The memristor was introduced in [1] as a new circuit element, but the words: "resists", "induces", "accumulates", "transforms", "amplifies", relate not only to electrical elements (*R, L, C*, etc.), , also to general human needs. A successful technical concept should obtain some use in everyday life. We do not have any good *uniting-all of the listed possibilities* word equivalent to "chooses". Perhaps, in the spirit of the switch application [2], "memrists" (e.g. as "memristing girl" wishing to become a bride, and never in the sense of "choosing the best moment") could be used.

ACKNOWLEDGMENT

I am grateful to Jean-Marc Ginoux for kind encouragement.

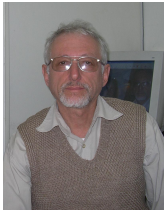 **Emanuel Gluskin** received the B.S. and M.S. degrees in physical engineering from Leningrad Polytechnic Institute (now Technical University) in 1974, and the Ph.D. degree in electrical engineering from Ben Gurion University, Beer Sheva, in 1990.

   He has industrial, research-authorities', and academic-teaching experiences.  He published in physics, electrical engineering, and mathematical journals, introducing some new concepts, for instance, the most relevant here "zerocrossing nonlinearity", a use of the inherently fractal nature of 1-ports, and a nonlinear transform, named "$\psi$-transform", for signal analysis.